\documentclass[aps,prl,showpacs,floatfix,twocolumn,byrevtex,superscriptaddress]{revtex4-1}
\usepackage{amsmath}
\usepackage{amssymb}
\usepackage{amstext}
\usepackage{amsopn}
\usepackage{amsfonts}
\usepackage{amsxtra}
\usepackage[english]{babel}
\usepackage{graphicx}
\usepackage{float}
\usepackage{bm}
\usepackage{multirow}
\usepackage{dcolumn}
\usepackage{hyperref}
\newcommand{\BKFA}{Ba$_{0.6}$K$_{0.4}$Fe$_{2}$As$_{2}$}

\newcommand{\BFAP}{BaFe$_{2}$(As$_{0.7}$P$_{0.3}$)$_{2}$}

\begin{document}

\title{\boldmath Hidden $T$-Linear Scattering Rate in Ba$_{0.6}$K$_{0.4}$Fe$_{2}$As$_{2}$ Revealed by
Optical Spectroscopy \unboldmath}
\author{Y. M. Dai}
\affiliation{Beijing National Laboratory for Condensed Matter Physics, Institute of Physics, Chinese Academy of Sciences, P.O. Box 603, Beijing 100190, China}
\affiliation{LPEM, ESPCI-ParisTech, CNRS, UPMC, 10 rue Vauquelin, F-75231 Paris Cedex 5, France}
\affiliation{Condensed Matter Physics and Materials Science Department, Brookhaven National Laboratory,
 Upton, New York 11973, USA}
\author{B. Xu}
\affiliation{Beijing National Laboratory for Condensed Matter Physics, Institute of Physics, Chinese Academy of Sciences, P.O. Box 603, Beijing 100190, China}
\affiliation{LPEM, ESPCI-ParisTech, CNRS, UPMC, 10 rue Vauquelin, F-75231 Paris Cedex 5, France}
\author{B. Shen}
\author{H. Xiao}
\affiliation{Beijing National Laboratory for Condensed Matter Physics, Institute of Physics, Chinese Academy of Sciences, P.O. Box 603, Beijing 100190, China}
\author{H. H. Wen}
\affiliation{Beijing National Laboratory for Condensed Matter Physics, Institute of Physics, Chinese Academy of Sciences, P.O. Box 603, Beijing 100190, China}
\affiliation{National Laboratory of Solid State Microstructures and Department of Physics,
Nanjing University, Nanjing 210093, China}
\author{X. G. Qiu}
\affiliation{Beijing National Laboratory for Condensed Matter Physics, Institute of Physics, Chinese Academy of Sciences, P.O. Box 603, Beijing 100190, China}
\author{C. C. Homes}
\email[]{homes@bnl.gov}
\affiliation{Condensed Matter Physics and Materials Science Department, Brookhaven National Laboratory,
 Upton, New York 11973, USA}
\author{R. P. S. M. Lobo}
\email[]{lobo@espci.fr}
\affiliation{LPEM, ESPCI-ParisTech, CNRS, UPMC, 10 rue Vauquelin, F-75231 Paris Cedex 5, France}

\date{\today}

%
%

\begin{abstract}
The optical properties of Ba$_{0.6}$K$_{0.4}$Fe$_{2}$As$_{2}$ have been determined in the normal state
for a number of temperatures over a wide frequency range. Two Drude terms, representing two groups of carriers with different scattering rates ($1/\tau$), well describe the real part of the optical conductivity, $\sigma_{1}(\omega)$. A ``broad'' Drude component results in an incoherent background with a $T$-independent $1/\tau_b$, while a ``narrow'' Drude component reveals
a $T$-linear $1/\tau_n$ resulting in a resistivity $\rho_n \equiv 1/\sigma_{1n}(\omega\rightarrow 0)$ also linear in temperature.
An arctan($T$) low-frequency spectral weight is also a strong evidence for a $T$-linear 1/$\tau$. Comparison to other materials with similar behavior suggests that the $T$-linear $1/\tau_n$ and $\rho_n$ in Ba$_{0.6}$K$_{0.4}$Fe$_{2}$As$_{2}$ originate from scattering from spin fluctuations and hence that an antiferromagnetic quantum critical point is likely to exist in the superconducting dome.
\end{abstract}



\maketitle

%
%

Over the last several decades, it has been observed that the electrical resistivity
$\rho$ of some strongly-correlated materials increases linearly with temperature ($T$-linear $\rho$),
particularly in the vicinity of an antiferromagnetic quantum critical point (QCP),
a striking deviation from Landau's Fermi-liquid description of metals.
This anomalous $T$-linear $\rho$, extensively studied in the high-$T_{c}$
cuprate superconductors \cite{Boebinger2009,Cooper2009, Daou2009,Jin2011}, organic Bechgaard
salts \cite{Doiron-Leyraud2009,Doiron-Leyraud2009a}, as well as heavy-fermion metals
\cite{Malinowski2005,Tanatar2007,Trovarelli2000,Gegenwart2002}, may be intimately
related to the emergence of superconductivity \cite{Jin2011,Taillefer2010,Doiron-Leyraud2009}.
It is generally believed that in proximity to the antiferromagnetic QCP, spin fluctuations are so strong that the scattering process of quasiparticles is severely modified, inducing non-Fermi-liquid behavior. Such spin fluctuations may be responsible for the pairing of electrons in high-$T_c$ superconductors \cite{Monthoux2007,Moriya2000,Mazin2008}.

The presence of the $T$-linear $\rho$ and a QCP in the newly discovered iron-based
superconductors is highly desired since superconductivity arises in the vicinity
of the spin-density-wave (SDW) instability \cite{Kamihara2008,Rotter2008}. Up to this point,
$T$-linear $\rho$ has been observed by transport, arousing considerable effort to describe
it as evidence of possible quantum criticality in iron-pnictides, especially in
the ``122'' family \cite{Xu2008,Gooch2009,Shen2011,Maiwald2012,Hashimoto2012,
Kasahara2010,Doiron-Leyraud2009,Shibauchi2013}.  However, unlike the high-$T_{c}$ cuprates,
iron-pnictides fall into the category of multi-band materials \cite{Singh2008,Ding2008}.  Up to five Fe-3$d$ bands crossing the Fermi level contribute to the Fermi surface, leading to the presence of three hole-like Fermi pockets at $\Gamma$-point
and two electron-like pockets at the corners of Brillouin zone. The scattering rate $1/\tau$
and the response of quasiparticles to external electrical field may vary considerably
in different Fermi pockets \cite{Fang2009,Shen2011,Muschler2009,Albenque2010}.
As a result, the transport properties become extremely complicated in such a system and the question of whether a $T$-linear $\rho$ in iron-pnictides originates from multi-band effects or the presence of a QCP makes it inadequate to investigate the transport properties alone in search of a
non-Fermi-liquid behavior and evidence for possible QCP \cite{Albenque2010,Albenque2009}.

Although many optical studies on iron-pnictides have been reported \cite{Li2008,Charnukha2011,Charnukha2011a}, the above issue has never been touched due to the absence of detailed $T$ dependent optical data. In this Letter we address this issue by studying the detailed $T$ dependence of the
optical conductivity and low-frequency spectral weight in Ba$_{0.6}$K$_{0.4}$Fe$_{2}$As$_{2}$.
The low-frequency optical conductivity is described by two Drude terms: a broad Drude with a large $1/\tau_b$ that is basically $T$-independent alongside a narrow one with a small $T$-linear $1/\tau_n$ that reveals a $\rho_n \equiv 1/\sigma_{1n}(\omega \rightarrow 0) \propto T$. The low-frequency spectral weight increases with cooling, following an $\arctan(T)$ dependence,
which is demonstrated to be a clear signature of $T$-linear $1/\tau$. Comparison with similar behavior found
in other materials attributes the $T$-linear $1/\tau_n$ and $\rho_n$
in Ba$_{0.6}$K$_{0.4}$Fe$_{2}$As$_{2}$ to spin fluctuation scattering and the presence
of a QCP in the superconducting dome.

%
%

\begin{figure}[tb]
\includegraphics[width=0.8\columnwidth]{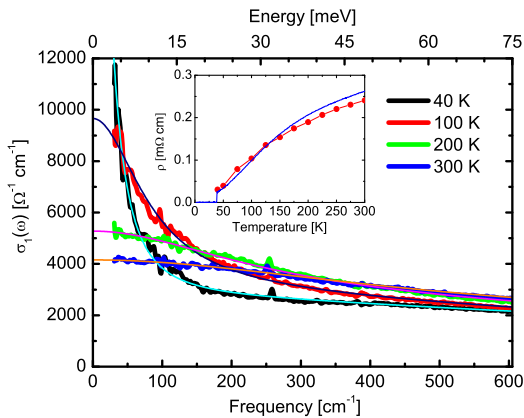}
\caption{ (color online) The thick solid lines in the main panel are {\em ab}-plane $\sigma_1(\omega)$ of
Ba$_{0.6}$K$_{0.4}$Fe$_{2}$As$_{2}$ at different temperatures in the
normal state. The smooth lines through the data are fits with Drude-Lorentz model. The inset
shows the DC resistivity as a function of temperature from transport measurement (solid
curve) and the values derived from the zero frequency extrapolation of $\sigma_1(\omega)$
(solid circles).}
\label{Sigma1}
\end{figure}

High quality Ba$_{0.6}$K$_{0.4}$Fe$_{2}$As$_{2}$ single crystals were
grown using a self-flux method \cite{Shen2011}. The inset of Fig.~\ref{Sigma1} shows the
DC resistivity of Ba$_{0.6}$K$_{0.4}$Fe$_{2}$As$_{2}$ as a function of temperature.
%
%
The $\rho(T)$ curve is characterized by a sharp superconducting transition at $T_c\simeq 39$~K
and a tendency to saturation at room temperature. A visible change in slope occurs at
about 175~K.

The near-normal incident reflectance, $R(\omega)$, has been measured for light polarized in the {\em a-b} plane using FTIR spectrometers and
an {\em in situ} evaporation technique \cite{Homes1993a}.  Data from $\simeq 20 - 12\,000$~cm$^{-1}$
were collected at 18 different temperatures from 5 to 300~K on a freshly cleaved surface. The visible and UV
range ($10\,000 - 55\,000$~cm$^{-1}$) $R(\omega)$ was measured at room temperature with an
AvaSpec-2048$\times 14$ fiber optic spectrometer.  The real part of the complex optical
conductivity, $\sigma_{1}(\omega)$, is determined from $R(\omega)$ via Kramers-Kronig analysis.
A Hagen-Rubens form ($R = 1 - A \sqrt{\omega}$) is used for the low-frequency extrapolation.
At high frequencies, $R(\omega)$ is assumed to be constant to 40~eV, above which a
free-electron response ($\omega^{-4}$) is used.

%
%

The main panel of Fig.~\ref{Sigma1} shows $\sigma_1(\omega)$ at
4 selected temperatures in the normal state (thick solid lines); all the spectra
exhibit the well-known Drude-like metallic response.  In order to quantitatively analyze
the optical data, we fit $\sigma_{1}(\omega)$ to the Drude-Lorentz model,
\begin{equation}
\sigma_{1}(\omega)=\frac{2\pi}{Z_{0}}\! \left[
  \sum_k\frac{\Omega^{2}_{p,k}}{\tau_k(\omega^{2}+\tau_k^{-2})}
  \!+\!
  \sum_j\frac{\gamma_j\omega^{2}\Omega_j^{2}}{(\omega_j^{2}-\omega^{2})^{2}+\gamma_j^{2}\omega^{2}}
  \right]
\label{DL}
\end{equation}
where $Z_{0}$ is the vacuum impedance. The first term describes a sum of free-carrier Drude
responses, each characterized by a plasma frequency
$\Omega_{p}^2 = 4\pi ne^2/m^\ast$, where $n$ is a carrier concentration and $m^\ast$ is an
effective mass, and a scattering rate $1/\tau$.  The second term corresponds to a sum of
Lorentz oscillators characterized by a resonance frequency ($\omega_j$), a line width
($\gamma_j$) and an oscillator strength ($\Omega_j$). This Drude-Lorentz model is also used to determine the dc properties of the system \cite{Romero1992}.

\begin{figure}[tb]
\includegraphics[width=0.8\columnwidth]{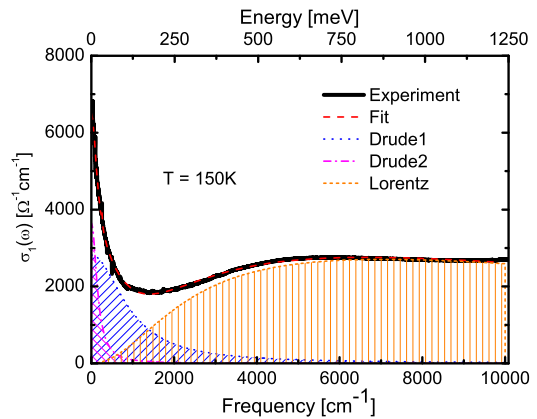}
\caption{ (color online) The black solid curve is the measured $\sigma_1(\omega)$ of
Ba$_{0.6}$K$_{0.4}$Fe$_{2}$As$_{2}$ at 150~K. The red dashed line through the data
is the fit which is decomposed into a broad Drude (blue dotted line), a narrow Drude
(pink dotted-dashed line) and a Lorentz (orange short-dashed line) term.}
\label{Fit}
\end{figure}

As shown in Fig.~\ref{Fit}, $\sigma_{1}(\omega)$ at 150~K is described by a broad Drude
with a large scattering rate $1/\tau_b \approx 936$~cm$^{-1}$, and a narrow Drude with
a small scattering rate $1/\tau_n \approx 158$~cm$^{-1}$ and an overdamped Lorentzian term.
%
%
The linear superposition of these three components gives very good description to
$\sigma_1(\omega)$ up to $10\,000$~cm$^{-1}$ at all the measured temperatures in the
normal state. Fits for other temperatures are selectively shown in the
main panel of Fig.~\ref{Sigma1} as smooth thin lines through the corresponding data.
The inset of Fig.~\ref{Sigma1} compares the optical estimate for the DC resistivity $\rho \equiv 1/\sigma_1(\omega \rightarrow 0)$ (solid circles) to the transport measurements (solid line).

The two-Drude fit indicates the existence of two groups of carriers with different
$1/\tau$'s in Ba$_{0.6}$K$_{0.4}$Fe$_{2}$As$_{2}$, which was first pointed out by Wu \emph{et al.}
in various iron-pnictides \cite{Wu2010a}.
The disparity of the $1/\tau$'s in different bands is also supported by both theoretical
calculations \cite{Kemper2011} and measurements using other techniques on similar
materials \cite{Fang2009,Shen2011,Albenque2010,Muschler2009}.
Tu {\em et al.} suggest that it is appropriate to describe the broad Drude term as bound excitations \cite{Tu2010}, because the mean free path $l = v_{F} \tau$ ($v_{F}$ is the Fermi velocity) associated with the broad Drude is less than the shortest interatomic spacing, violating the Mott-Ioffe-Regel condition \cite{Gurvitch1981}.
In Ba$_{0.6}$K$_{0.4}$Fe$_{2}$As$_{2}$, the average Fermi velocities of the electron and hole pockets
are estimated to be $v^{e}_{F} \simeq 0.40$~eV{\AA} and $v^{h}_{F} \simeq 0.36$~eV{\AA}
\cite{Ding2011}, respectively.  Furthermore, it is reported that in iron-pnictides holes have a larger $1/\tau$ than electrons \cite{Fang2009,Shen2011,Albenque2010,Muschler2009}.
If we associate the broad Drude component ($1/\tau_b \simeq 936$~cm$^{-1}$)
with the hole pockets, a mean free path of $l_{h} \simeq 3$~{\AA} is obtained. This value is
close to the lattice parameter $a \simeq 4$~{\AA} of the 122 family compounds and probably
too small for coherent transport. Since the broad Drude only produces an incoherent, $T$-independent, background contribution to the total $\sigma_{1}(\omega)$, the nature of the broad Drude component (whether or not bound excitations) does not affect our analysis
on the temperature dependence of $\sigma_1(\omega)$ and low-frequency spectral weight.

%
%
\begin{figure}[tb]
\includegraphics[width=0.9\columnwidth]{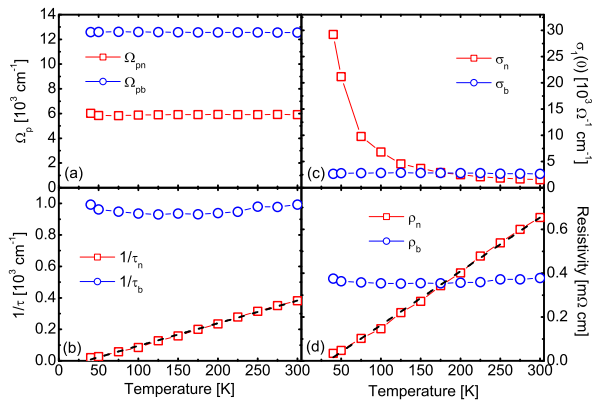}
\caption{ (color online) The $T$ dependence of (a) the plasma frequency $\Omega_p$, (b) the scattering rate $1/\tau$, (c) the contribution to DC conductivity $\sigma_1(\omega \rightarrow 0)$, and (d) the equivalent resistivity $\rho$ for the narrow and broad Drude components, respectively. The dashed lines in panels (b) and (d) are linear fits to $1/\tau_n$ and $\rho_n$, respectively.}
\label{FitPar}
\end{figure}
The temperature dependence of the two Drude components provides information about the nature of
the two different types of carriers in this material. Figure~\ref{FitPar} shows the
$T$ dependence of the Drude parameters from our fits. The subscripts n and b stand for the narrow and broad Drude terms,
respectively. Fig.~\ref{FitPar}(a) shows the $T$ dependence of the plasma frequencies
of the two Drude terms. Upon cooling, $\Omega_p$ is roughly a constant for each of the two
components, indicating that the band structure and $n/m^\ast$ does not change with temperature,
in agreement with a previous work \cite{Wu2010a}.
Fig.~\ref{FitPar}(b) portrays the $T$ dependence of the scattering rate of the
two Drude components, where $1/\tau_b$ is basically $T$-independent while $1/\tau_n
\propto T$; the black dashed line denotes a linear fit.
Fig.~\ref{FitPar}(c) displays the contribution of the two groups of carries to the DC conductivity.
As the broad Drude ($\sigma_b$) is $T$-independent, the temperature dependence of the total DC conductivity arises
out of the narrow Drude band ($\sigma_n$). The $\rho(T)$ curve of Ba$_{0.6}$K$_{0.4}$Fe$_{2}$As$_{2}$,
shown in the inset of Fig.~\ref{Sigma1}, exhibits a tendency to saturation at room temperature, and a
change of slope can be seen at about 175~K.  This behavior can be explained by the different $T$
dependence of the two Drude bands, which can be considered as a parallel-circuit \cite{Wiesmann1977}:
$\sigma = \sigma_n + \sigma_b$.
A crossover-region, where $\sigma_n \simeq \sigma_b$, is found in Fig.~\ref{FitPar}(c) at $\simeq 175$~K.
Below this temperature, $\sigma_n > \sigma_b$, so the total DC conductivity is dominated by $\sigma_n$
which exhibits strong temperature dependence. As a result, below 175~K, $\rho(T)$ decreases
quickly with decreasing temperature.
Above 175~K, $\sigma_n < \sigma_b$,  and the total DC conductivity is dominated by $\sigma_b$ which shows
no temperature dependence. Hence, above 175~K, the growth of the DC resistivity slows with heating,
resulting in the change of slope and the tendency to saturation in $\rho(T)$.
Similar conclusions were obtained from investigations of the Hall effect \cite{Albenque2010,Albenque2009} and theoretical calculations \cite{Golubov2011}.
In the $\omega \rightarrow 0$ limit, the inverse of $\sigma_{1}(0)$
yields the resistivity from the two Drude components, as shown in Fig.~\ref{FitPar}(d). The resistivity
of the broad Drude remains a constant at all measured temperatures, while a $T$-linear $\rho_n$ is
revealed for the narrow Drude component.  This is in accord with transport measurements on hole-doped
122 compounds \cite{Shen2011,Maiwald2012,Gooch2009}, where $T$-linear $\rho$ was observed at low temperatures in optimally-doped materials. The $T$-linear $\rho$ is only found at low temperatures as this is the region dominated by the narrow Drude component.

Further evidence for $T$-linear $1/\tau$ lies in the temperature dependence of the low-frequency
spectral weight. The spectral weight is defined as
\begin{equation}
  W_{0}^{\omega_{c}} = \int_{0}^{\omega_{c}} \sigma_{1}(\omega) d\omega ,
\label{SW}
\end{equation}
where $\omega_{c}$ is a cutoff frequency.
In a Drude metal the scattering rate decreases upon cooling, producing a narrowing of
the Drude response and an increase of the DC conductivity, resulting in a transfer of spectral
weight from high to low frequencies and an increase in $W_{0}^{\omega_{c}}$.
To quantitatively analyze the $T$ dependence of the low-frequency spectral weight, we adopt one Drude optical
conductivity [see Eq.~(\ref{DL})] into Eq.~(\ref{SW}), to obtain the spectral weight as a function of $1/\tau$
\begin{equation}
 W_{0}^{\omega_{c}}(1/\tau) = \frac{2 \pi \Omega_{p}^{2}}{Z_{0}}
   \arctan \left( \omega_{c} \tau \right).
\label{SWDrude}
\end{equation}
In the case of $1/\tau \propto T$, Eq.~(\ref{SWDrude}) can be simplified as
\begin{equation}
  W_{0}^{\omega_{c}}(T) = a_{1} \arctan \left( \frac{a_{2}}{T} \right),
\label{SWDrudevsT}
\end{equation}
where $a_{1} = {2 \pi \Omega_{p}^{2}}/Z_{0}$, and $a_{2} \propto \omega_c$; both are $T$-independent
parameters. Considering the spectral weight arising from the incoherent part and inter-band transition (Lorentz), which are both $T$-independent, we introduce the third $T$-independent parameter
$a_{0}$ into Eq.~(\ref{SWDrudevsT}).  Finally, the low-frequency spectral weight as a function of temperature
for $1/\tau \propto T$ is written as
\begin{equation}
  W_{0}^{\omega_{c}}(T) = a_{0} + a_{1} \arctan (\frac{a_{2}}{T}).
\label{SWvsT}
\end{equation}

\begin{figure}[tb]
\includegraphics[width=0.8\columnwidth]{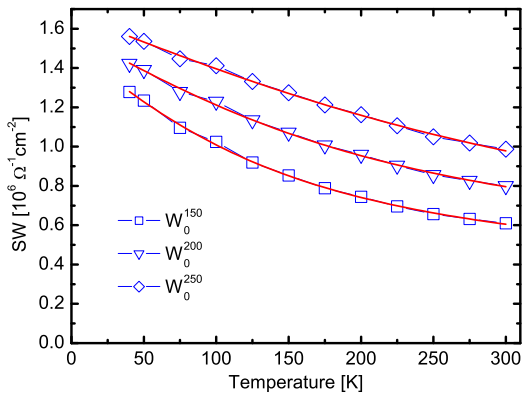}
\caption{ (color online) Temperature dependence of the low-frequency spectral weight
$W_{0}^{\omega_c}$ with different $\omega_c$'s: 150 (squares), 200 (triangles) and
250~cm$^{-1}$ (diamonds). The solid lines through the data represent the expected
temperature dependence for $1/\tau \propto T$ at each $\omega_c$.}
\label{FigRSW}
\end{figure}

$W_{0}^{\omega_c}$ can be easily determined by integrating the measured $\sigma_{1}(\omega)$.
The open symbols in Fig.~\ref{FigRSW} denote $W_{0}^{\omega_{c}}$
with different $\omega_c$'s: 150 (squares), 200 (triangles) and 250~cm$^{-1}$ (diamonds).
As expected in a metallic system, for all the $\omega_c$'s, $W_{0}^{\omega_{c}}$ increases
with decreasing $T$. The solid curves through the data are the least square fits
using Eq.~(\ref{SWvsT}).  The excellent agreement indicates that a $T$-linear $1/\tau$ dominates
the low frequency $\sigma_1(\omega)$ over a very broad temperature range.

We now discuss the possible origin of the $T$-linear $1/\tau_n$ and $\rho_n$.
A $T$-linear $\rho$ is usually expected in a system dominated by electron-phonon scattering
in the regime $T \gg \Theta_{D}$, where $\Theta_{D}$ is the Debye temperature; in the regime
$T < \Theta_{D}$, the resistivity is approximated by a low-order polynomial that is neither
linear nor quadratic in temperature. Transport measurements in the 122 compounds suggest
$\Theta_D \simeq 250$~K \cite{Ni2008,Tu2010}.  However, the $T$-linear $1/\tau_n$ and $\rho_n$ spans a
very broad temperature range, from $T_{c}$ up to 300~K, which is inconsistent with
the electron-phonon dominated scattering process.  Further clues on the $T$-linear $1/\tau_n$
and $\rho_n$ may be revealed by an examination of the phase diagram of the
Ba$_{1-x}$K$_{x}$Fe$_{2}$As$_{2}$ system and a comparison with similar behavior found in other
materials.
The parent compound BaFe$_2$As$_2$ orders in an SDW state below $T_{SDW} \approx 138$~K; $T_{SDW}$
is suppressed by K doping allowing superconductivity to emerge.  Here, $1/\tau_n \propto T$
is observed at the doping where the SDW order is entirely suppressed, i.e. $T_{SDW} \rightarrow 0$.
At this point, spin fluctuations are expected to be very strong, which has been experimentally
confirmed by NMR \cite{Ning2010,Nakai2010,Li2011}.
This brings us to models that attribute $T$-linear $\rho$ to spin-fluctuation
scattering \cite{Moriya2000,Sachdev2011}.  The $T$-linear $\rho$ (or $1/\tau$) is also
found in cuprates such as Nd-doped La$_{2-x}$Sr$_x$CuO$_4$ \cite{Daou2009} and electron-doped
La$_{2-x}$Ce$_x$CuO$_4$ \cite{Jin2011}, organic superconductors (TMTSF)$_{2}$PF$_{6}$
\cite{Doiron-Leyraud2009} and (TMTSF)$_{2}$ClO$_{4}$ \cite{Doiron-Leyraud2009a}, as well as a
number of heavy fermions such as CeCoIn$_{5}$ \cite{Malinowski2005,Tanatar2007} and YbRh$_{2}$Si$_{2}$
\cite{Trovarelli2000,Gegenwart2002}.  Studies on these materials have shown that $T$-linear $\rho$,
arising on the border of antiferromagnetic order, is caused by spin-fluctuation scattering due
to the proximity of an antiferromagnetic QCP.  Studies on cuprates and Bechgaard salts
\cite{Taillefer2010,Jin2011} further show that the strength (or coefficient) of the $T$-linear
$\rho$ scales with $T_{c}$ and disappears upon approaching the point where $T_{c} \rightarrow 0$,
suggestive of intimate relation between the $T$-linear $\rho$ and superconductivity.
Anomalous $T$-linear $\rho$ (or $1/\tau$) and pairing in unconventional superconductors may share a
common origin: spin fluctuations.  The hidden $T$-linear $1/\tau_n$ and $\rho_n$
in Ba$_{0.6}$K$_{0.4}$Fe$_{2}$As$_{2}$ revealed by optical measurements may have the same origin as
those found in the cuprates, organic superconductors and heavy fermion metals, because these materials
share strikingly similar phase diagrams. Our observations may also imply a possible QCP in the
superconducting dome. The existence of the QCP in iron-pnictides is supported by transport properties \cite{Gooch2009,Shen2011,Maiwald2012}, NMR studies \cite{Ning2010,Nakai2010}, de Haas-van Alphen effect \cite{Walmsley2013}, penetration depth measurement \cite{Hashimoto2012} and first-principles calculations \cite{Xu2008}.

Spin-fluctuation induced $T$-linear $\rho$ suggests an equivalent $\omega$-linear $1/\tau(\omega)$ \cite{Moriya2000}, obtainable through the extended Drude model provided that interband contribution is negligible. Low energy interband transitions are important in iron-pnictides, and their contribution has to be subtracted to determine $1/\tau(\omega)$ for mobile carriers \cite{Benfatto2011,Charnukha2011a}. We calculated $1/\tau(\omega)$ via the extended Drude model with the interband contribution subtracted (supplementary) and found that by taking into account the interband transitions, a large fraction of the frequency dependence in $1/\tau(\omega)$ is eliminated, which is consistent with \citeauthor{Charnukha2011a}'s analysis on Ba$_{0.68}$K$_{0.32}$Fe$_2$As$_2$ \cite{Charnukha2011a}. There is no confident evidence for the expected $\omega$-linear $1/\tau(\omega)$, since it could be masked by the multiband character of the iron-pnictides.

In order to check if the $T$-linear $1/\tau$ is unique in \BKFA\ or general in iron-pnictides, we applied the same analysis to \BFAP\ (supplementary). Interestingly, $T$-linear $1/\tau$ is also found for the narrow Drude. This suggests that $T$-linear $1/\tau$ is not unique in \BKFA, but most likely, a general behavior in iron-pnictides at the doping where SDW order is completely suppressed, i.e. $T_{SDW} \rightarrow 0$.

%
%
In summary, the detailed $T$ dependence of the normal state $\sigma_1(\omega)$
and low-frequency spectral weight in Ba$_{0.6}$K$_{0.4}$Fe$_{2}$As$_{2}$
have been examined. Two Drude components with different $1/\tau$'s yield an excellent
description of the low-frequency optical response, indicating the existence of two groups
of carriers with different quasiparticle lifetimes.  The broad Drude component produces
an incoherent background conductivity with no temperature dependence, while the narrow Drude
component reveals a $T$-linear $1/\tau_n$ and $\rho_n$.  This fact explains the
$T$-linear $\rho$ behavior at low temperatures and the tendency to saturation at
room temperature observed in transport measurements in optimally hole-doped 122 compounds.
An arctan($T$) dependence of the low-frequency spectral weight is also a strong evidence
for a $T$-linear $1/\tau$. Comparison with similar behavior found in other
materials suggests that the $T$-linear $1/\tau_n$ and $\rho_n$ in Ba$_{0.6}$K$_{0.4}$Fe$_{2}$As$_{2}$ may arise
out of spin-fluctuation scattering due to the possible existence of an antiferromagnetic QCP
in the superconducting dome.

%
%

We thank Hu Miao, Xiaoxiang Xi, Wei Ku, Cong Ren and Lei Shan for helpful discussion.
Work in Beijing was supported by the NSFC (No. 91121004 and
No. 11104335) and the MSTC (973 Projects No. 2011CBA00107,
No. 2012CB821400, No. 2012CB921302 and No. 2009CB929102). Work at BNL was supported by the U.S.
Department of Energy, Office of Basic Energy Sciences, Division of Materials Sciences and
Engineering under Contract No. DE-AC02-98CH10886. We acknowledge the financial support from
the Science and Technology Service of the French Embassy in China.

%
%


%

\end{document}